\documentclass[aps,prd, 
               preprintnumbers,numbers,sort&compress,nofootinbib,
               superscriptaddress,showpacs,
               colorlinks,
               linkcolor=blue,
               citecolor=blue]{revtex4-1}

\usepackage{amsmath}
\usepackage{amssymb}
\usepackage{hyperref}

\usepackage{graphicx,amsmath,amssymb,bm}
\usepackage{psfrag}
\newcommand{\exclude}[1]{}

\newcommand{\beq}{\begin{equation}}
\newcommand{\eeq}{\end{equation}}
\newcommand{\bea}{\begin{eqnarray}}
\newcommand{\eea}{\end{eqnarray}}

\def\la{\langle }
\def\ra{ \rangle }
 \newcommand{\junk}[1]{}

\begin{document}

\title{Conformal window  in
QCD for  large numbers of colours and flavours} 
 
  \author{Ariel R. Zhitnitsky}
\affiliation{Department of Physics and Astronomy, University of
  British Columbia, Vancouver,  Canada}
 %\date{\today}

\begin{abstract}
We conjecture that  the   phase transitions in QCD at large number of colours $N\gg 1$  is triggered by the  drastic change in the instanton density. As a result of it, all physical observables   also experience some sharp modification in  the $\theta$ behaviour. This conjecture is motivated by   the holographic model of QCD  where 
confinement -deconfinement phase transition 
 indeed happens precisely at   temperature $T=T_c$ where $\theta$ dependence of the vacuum energy
experiences a sudden change in behaviour: from $N^2\cos(\theta/N)$ at  $T<T_c$ to $\cos\theta\exp(-N)$ at $T>T_c$.  This conjecture is also supported by recent lattice studies. We employ this conjecture to study  a possible phase transition  as a function 
of $\kappa\equiv N_f/N$ from confinement  to conformal phase  in the Veneziano limit $N_f\sim N$
when  number of flavours and colours are large, but the ratio   $\kappa$  is finite. 
  Technically, we consider an operator which
gets its expectation value solely from nonperturbative instaton effects.
When $\kappa$ exceeds some critical value $\kappa> \kappa_c$
 the integral over instanton size is dominated by small-size
instatons, making the instanton computations  reliable with expected  $\exp(-N)$ behaviour.  However,
when $\kappa<\kappa_c$, the integral over instaton
size is dominated by large-size instantons, and the instanton expansion breaks down. 
 This regime with $\kappa<\kappa_c$   corresponds to the confinement phase.  We also compute the variation  of the critical $\kappa_c( T, \mu)$  when the temperature and chemical potential $T, \mu \ll \Lambda_{QCD}$ slightly vary. 
 We also discuss the  scaling  $(x_i-x_j)^{-\gamma_{\rm det }}$ in the conformal phase.
\end{abstract}
 
\pacs{ }
\maketitle

\section{ Introduction}
Understanding the phase diagram at nonzero external parameters $T,
\mu, \kappa $ and the so-called $\theta$-parameter 
is one of the
most hard problem in QCD. Obviously, this area is a prerogative
of numerical lattice computations.  However, some insights about the
basic features of the phase diagram may be inferred by using some
analytical methods. Our  study of this hard question with non-zero $\theta$ parameter  has been motivated, in fact,     by  very deep  cosmological connections  related to  the QCD phase transition in early universe when $\theta$ parameter is thought  to be nonzero. However, the cosmological connections shall not be elaborated in the present work\footnote{\label{CP}Non-vanishing $\theta\neq 0$   implies that  the ${\cal{CP}}$ symmetry is strongly broken  during  the cosmological QCD phase transition. At the same time, a non-observation of any ${\cal{CP}}$ violating processes in strong interactions at the present epoch implies that the  $\theta$ parameter vanishes now.
  The well known (and generally accepted) resolution of the  strong ${\cal{CP}}$ problem is provided by the dynamical   axion field, which, at the same time,  might be a natural dark matter candidate. In different words, a unique and strong  source of the   ${\cal{CP}}$  violation which was available   during the QCD phase transition is not available anymore at present time. This unique source of the strong ${\cal{CP}}$ violation might be a missing ingredient  for   understanding  
 the  observed   the baryon- antibaryon asymmetry existing in nature, see few comments and references on cosmological connection  in Conclusion in section \ref{conclusion}.}.

The  approach we advocate in the present work  to attack this hard problem is based on a conjecture originally formulated in refs. \cite{Parnachev:2008fy,Zhitnitsky:2008ha}    that the de-confined phase transition is always accompanied by  very sharp changes in $\theta$ behaviour. Therefore, in principle, if our conjecture is correct, one can use any order parameter which nontrivially depends on $\theta$ and study this dependence on two sides of the phase transition line.  Very natural question immediately emerges: why and how these two apparently  very different things
(phase transition vs sharp $\theta$ changes) could be linked? What is the basic motivation for this proposal? First of all, this criteria is motivated by   the observation that in holographic model of QCD  the 
confinement -deconfinement phase transition indeed 
 happens precisely at the temperature $T=T_c$ where $\theta$ dependence 
experiences a sudden change in behaviour, see \cite{Parnachev:2008fy,Gorsky:2009me} and many related references therein. 

Secondly,  the proposal  is supported by  the numerical lattice results \cite{Alles:1996nm} -\cite{ Bonati:2013dza}, see also a review article \cite{Vicari:2008jw},  which unambiguously
suggest that the topological fluctuations related to $\theta$ (in particular the topological susceptibility) are strongly suppressed in deconfined phase, and this suppression becomes more severe with increasing $N$.

Thirdly, our new criteria is based on a physical picture which can be
shortly summarized as follows.  On one side of the phase transition
line the instanton gas is dilute with density $\sim e^{-N\gamma(\kappa,
\mu, T)}$ which implies a strong suppression of the topological
fluctuations at large $N$ with $\gamma(\kappa, \mu, T)>0$, see below
some details on generic features  of $\gamma(\kappa, \mu, T)-$ function. The calculations in
this region are under complete   theoretical control and the
vacuum energy has a nice analytic behaviour $\sim \cos\theta
e^{-N\gamma(\kappa, \mu, T)}$ as function of $\theta$.  At the
critical value   $\gamma(\kappa, \mu, T)$
changes the sign, the instanton expansion suddenly breaks down (at very large $N$), and one should
naturally expect that   there must be a
sharp transition in $\theta$ behavior as simple formula $\sim
\cos\theta$ can only be valid when the instanton gas is dilute and
semiclassical calculation is justified, which is obviously not the
case  when the instanton density formally becomes exponentially large $\sim e^N$.  
This is a region of confinement
when  the dilute instanton gas approximation can not be trusted anymore.  What happens to the
well-defined objects (instantons) as the phase transition line is crossed
at  the critical values?  One can argue that the instantons do not
completely disappear from the system,  but rather dissociate into the instanton
quarks\footnote{\label{constituents}Instanton quarks, also known as ``fractional instantons" or ``instanton partons",   originally appeared in 2d models.
Namely, using an exact accounting and re-summation of the $n$-instanton
solutions in 2d CP$^{N-1}$ models, the original problem of a
statistical instanton ensemble was mapped unto a 2d Coulomb Gas (CG)
system of pseudo-particles with fractional topological charges $\sim
1/N$ \cite{Fateev}.  This picture leads to the elegant explanation of
the confinement phase and other important properties of the
$2d~CP^{N-1}$ models \cite{Fateev}.   We use   term the ``instanton quarks" to  emphasizes that there are precisely $N$ constituents making an instanton, similar to $N$ quarks making a baryon. Other suggested terms such as  the ``fractional instantons" or the ``instanton partons" do not quite reflect this
unique $1/N$ feature. These objects do not appear individually in path integral; instead, they appear as configurations consisting $N$ different   objects with fractional charge $1/N$ such that the total topological charge of each configuration   is always  integer. In this case $4Nk$ zero modes for $k$ instanton solution is interpreted as $4$   translation zero modes modes accompanied by  every single instanton quark. The same counting holds, in fact,  for any gauge group $G$, not limited to $SU(N)$ case. 
While the instanton quarks emerge in the path integral coherently, 
  these objects are highly delocalized: they may emerge on opposite sides of the space time or be close to each other with alike   probabilities. 
Similar objects have been discussed in a number of papers in a different context, see e.g. \cite{Belavin,vanbaal,Diakonov:2004jn,Zhitnitsky:2006sr,Collie:2009iz,Bolognesi:2011nh}. In particular, it has been argued that the well-established $\theta/N$ dependence in confined phase unambiguously implies that the relevant configurations in confined QCD must carry   fractional topological charges, see   \cite{Parnachev:2008fy,Zhitnitsky:2006sr} and references on original papers therein. The weakly coupled deformed QCD model \cite{Shifman:2008ja,Unsal:2008ch,Poppitz:2009uq,Poppitz:2009tw, Thomas:2011ee,Unsal:2012zj,Poppitz:2012nz,Anber:2013sga,Poppitz:2013zqa},   to be discussed in the next paragraph,  is a precise dynamical realization of this idea when the fractionally charged constituents play the dominant role, and  when
  the question on dissociation of the instantons to the instanton quarks can be explicitly tested and studied in the weak coupling regime.}, the objects with fractional
topological charges $\pm 1/N$ which become the dominant
pseudo-particles in the confined phase, which  is precisely the conjecture advocated in refs. \cite{Parnachev:2008fy,Zhitnitsky:2008ha}. 
\newpage

Finally, this entire framework can be in principle tested using some deformations of QCD, which on the one hand 
 preserve all the crucial elements of strongly interacting QCD, including confinement, nontrivial $\theta$ dependence, degeneracy of the topological sectors,  chiral symmetry breaking, etc. On the other hand the deformations are designed in such a way that they bring the system into the    weak coupling regime when all  computations are under complete theoretical control. 
In fact, the corresponding technique is well developed by now,  see  relevant for present studies references  \cite{Shifman:2008ja,Unsal:2008ch,Poppitz:2009uq,Poppitz:2009tw, Thomas:2011ee,Unsal:2012zj,Poppitz:2012nz,Anber:2013sga,Poppitz:2013zqa}. These recent studies essentially support the basic picture that the phase transition occurs as a result of complete reconstruction of the dominant  pseudoparticles on two sides of the phase transition line (instantons versus instanton quarks).   The system experiences a sharp transition in $\theta$ behaviour precisely as  a result of this reconstruction of the dominant pseudoparticles.   In other words, the driving force of a deconfined phase transition is the dissociation of the instantons into their constituents, the instanton quarks. This reconstruction obviously leads to the drastic changes in $\theta$ behaviour  on two sides of the phase transition line. We use this sharp alteration   in $\theta$ behaviour as a signal of the phase transition. It  precisely represents the basic conjecture formulated in refs. \cite{Parnachev:2008fy,Zhitnitsky:2008ha}.

   Therefore, the phase transition within this framework can be interpreted (with some very important reservations, see below) as Berezinskii-Kosterlitz-Thouless (BKT) -like phase transition\cite{BKT}:  at $T> T_{c}$
   (or $\mu>\mu_c$, or $\kappa >\kappa_c$ depending  on a specific  slice of the multidimensional phase diagram   we are interested in)
   the   constituents prefer to organize a single  instanton of a finite size. We coin this phase as a  ``molecular phase" which corresponds to a de-confined phase
   in conventional terminology. When one crosses the  phase transition line  at $T< T_{c}$
   (or $\mu<\mu_c$ or $\kappa <\kappa_c$)   the constituents prefer to stay far away 
     from each other.
   It corresponds to the dissociation of each instanton into $N$ constituents, and we 
   call this state  as the   ``$N$ component plasma phase" in 4d Euclidean space. This regime corresponds to  the confined phase
   in conventional terminology when all instanton quarks are delocalized in 4d Euclidean space. The gap in this confined phase is determined by the Debye correlation length of this 4d   plasma\footnote{\label{BKT}The BKT picture normally suggests that the correlation length expressed in terms of parameter $(\kappa-\kappa_c)$
   should exhibit a very specific behaviour when $(\kappa-\kappa_c)$
 crosses the phase transition line. To be more specific,  in ``molecular phase" the correlation length $\zeta$ should be infinite, while in the ``plasma phase" for $(\kappa_c-\kappa) >0$ it should demonstrate a specific BKT scaling $\zeta\sim  \exp(\frac{1}{\sqrt{\kappa_c-\kappa}})$ in close vicinity of the critical point. Unfortunately, our approach breaks down in the ``plasma phase" as we review in next section \ref{review}. Therefore,
  while the physical description of the phase transition in terms of the pseudo-particles is very similar to BKT picture,  the  corresponding analysis of the correlation length $\zeta$ in the vicinity of the phase transition    with   $(\kappa_c-\kappa) >0$ can not be carried out within this approach, see also a comment on this subject in concluding section \ref{conclusion}.}. We should comment here that the idea on possibility of the BKT type transition in case when parameter $\kappa$ varies was suggested previously~\cite{Kaplan:2009kr}, though the arguments and motivation of ref.\cite{Kaplan:2009kr} were very different from those advocated in the present work. 
  
  To avoid any confusion  in our future discussions, we should emphasize from the very beginning   that this analogy should  not be taken literally. Indeed, we have pseudo-particles which live in the 4d Euclidean space, 
 instead of real particles/quasi-particles in the original BKT picture.    The pseudo-particles  in 4d space have an interpretation  of  an auxiliary objects which  describe the tunnelling events,
  rather than static object which live in real Minkowski space time.  Furthermore, our term the ``molecular phase" 
  which corresponds to a de-confined phase in conventional terminology should not be confused with  molecular phase  in the original BKT picture when real particles/quasi-particles make a static bound state. In 4d Euclidean space the ``molecular phase"  implies that the corresponding Euclidean configurations (instantons) provide  the dominant contribution into the path integral. The same comment also applies to the ``$N$ component plasma phase" in 4d Euclidean space, which should not be confused with plasma phase  in the original BKT picture. This comment obviously implies that conventional entropy arguments can not be applied to 4d  ``phases" as the merely notion of the entropy does not exist  for such Euclidean objects. 
  Nevertheless, this analogy could be quite useful in the description of the universal properties of the phase transitions due to many formal similarities between these two   (very different) systems as we discuss in this work. Furthermore, from the 5d (holographic) view point  
  the 4d Euclidean objects (instantons and the instanton quarks) can be indeed interpreted as some static objects. However, we shall not elaborate on this holographic  interpretation  in the present work. 
  
  \begin{figure}[t]
\begin{center}
\includegraphics[width = 0.5\textwidth]{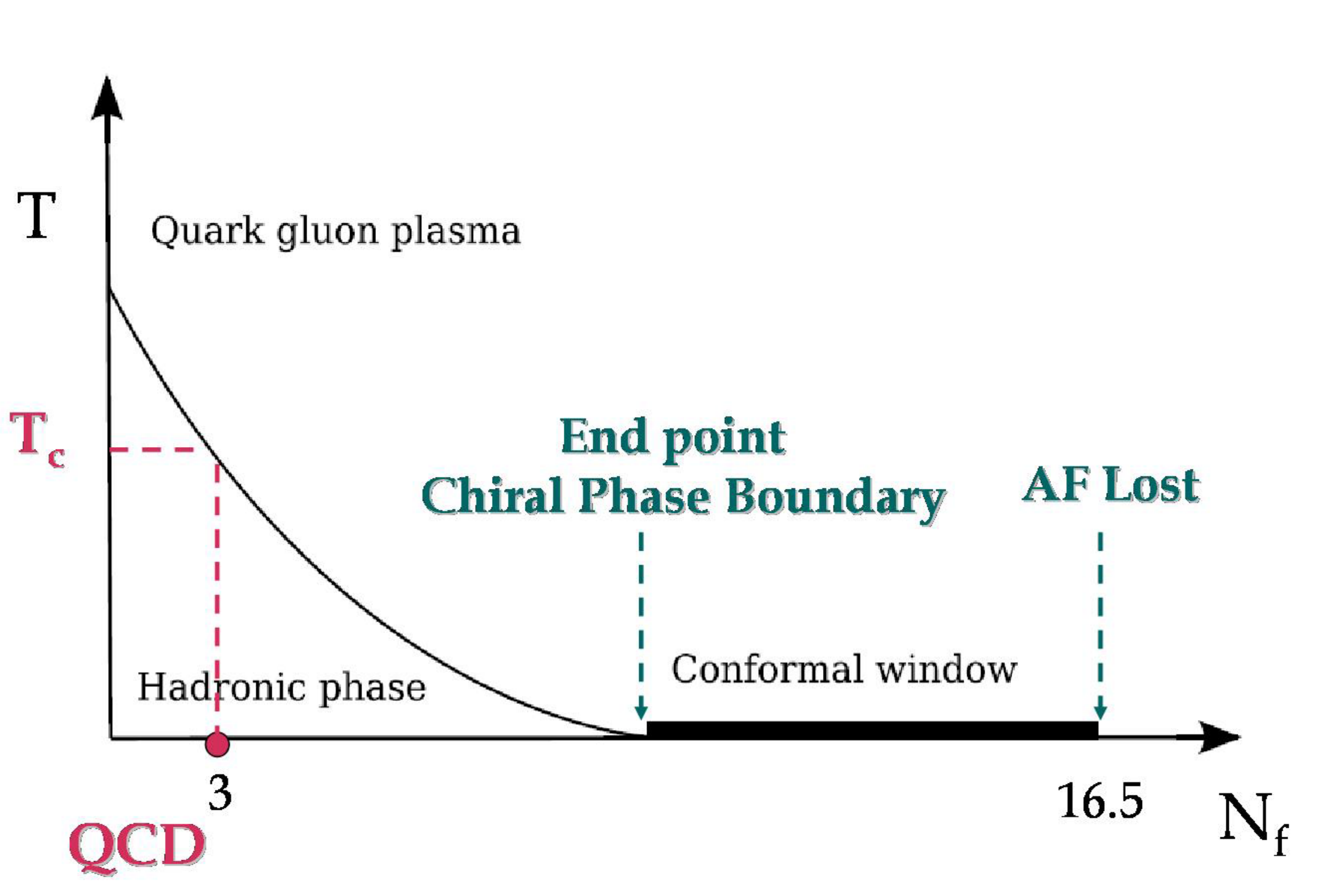}
\caption{ Assumed  phase diagram as a function of temperature $T$ and  number of flavours $N_f$,  taken from\cite{Pallante:2009hu}. }
\label{phase}
\end{center}
\end{figure}

To reiterate: we identify  the sharp changes in $\theta$ behaviour  in pure gauge theory with some kind  of BKT- like phase transition
 when in one phase the instanton constituents prefer  to be in ``molecular phase" while in another phase  they prefer to be in the ``plasma phase".
 In particular, in case with  variation of temperature  $T$  the corresponding studies
 have been carried out  in 
 \cite{Parnachev:2008fy} while  the case when    the chemical potential $\mu$ varies, the corresponding 
 results  have been  presented   in~\cite{Zhitnitsky:2008ha}.  The goal of this work  is to study confinement- deconfinement phase transition with variation of the   parameter 
$\kappa$.  We argue that the confined phase ceases to exist for sufficiently large $\kappa>\kappa_c$
as a result of reconstruction of the dominant pseudopartciles, similar to our studies 
of the phase transition lines at $T>T_c$ or $\mu>\mu_c$. It is normally  assumed that  the asymptotic freedom still holds in this  region as plotted on Fig. \ref{phase}.    At zero temperature the confined phase is  expected to be replaced by a ``non- Abelian Coulomb phase", conventionally referred to as  the ``conformal window", see   some original analytical results \cite{BZ, Miransky:1996pd,Appelquist:1996dq, Appelquist:1998rb,Sannino:1999qe,Dietrich:2006cm,Ryttov:2007sr,Pica:2010mt},  the recent lattice  computations\cite{Appelquist:2007hu,Appelquist:2009ty,Deuzeman:2009mh,Fodor:2011tu,Appelquist:2011dp,Deuzeman:2013kma} and  review \cite{Pallante:2009hu}
 on the subject. 
 
 To accomplish our goal  we use $\la {\rm det }~ \bar{\psi}_L^f
\psi_R^f \ra$ as  the   order parameter  to analyze the phase transition as function of
$\kappa$.  This operator   breaks U(1)$_A$
symmetry, and gets its expectation value from instantons.   The variation  of this potential  as a function of $\kappa$ detects  the   drastic  reorganization   of the dominant 
pseudoparticles (instanton vs instanton quarks).   Therefore, we study the behaviour of the expectation value  $\la {\rm det }~ \bar{\psi}_L^f
\psi_R^f \ra$ at  $\kappa>\kappa_c$ in the ``molecular phase" where the instanton expansion is under complete theoretical control. We slowly approach the regime where the instanton expansion breaks down.
  We identify this point with  
the critical point $\kappa_c$.  We argue that in large $N$ limit we can   approach this point with arbitrary high precision as the instanton density remains to be exponentially suppressed (and therefore instanton expansion remains to be justified) as long as inequality  $ (\kappa-\kappa_c)\gg  {1}/{N}$ holds. 
We use the same guiding principles formulated in  \cite{Parnachev:2008fy, Zhitnitsky:2008ha} to   estimate how this
critical value $\kappa_c$ depends on temperature and chemical potential
$\kappa_{c}(T, \mu)$ for  sufficiently small $T, \mu$. This analysis  essentially makes a solid prediction (of course, within our framework) on how the conformal window extends to  non-vanishing temperature and chemical potential, 
which is the subject of  section \ref{conformal}. Our presentation starts with section \ref{review}  where we review 
   our basic conjecture and guiding principles originally formulated  in refs.  \cite{Parnachev:2008fy, Zhitnitsky:2008ha}.
   Section \ref{conclusion} is our conclusion   with  few comments on   future directions and possible tests of the  entire framework.

 \section{Confinement- Deconfinement  Phase Transition in hot  and dense QCD at large $N$.}\label{review}
 
 We start with a short review of ref.\cite{Parnachev:2008fy} where the conjecture 
 (that  the confinement-deconfinement phase transition  happens precisely where $\theta$ behavior 
 sharply changes) was implemented for large $N$ QCD at $T\neq 0$.
 Such a sharp transition is indeed observed  in the holographic model of QCD.  From quantum field theory viewpoint such a transition 
 can be understood   as follows. Instanton calculations are under complete theoretical 
 control in the region $T>T_c$ as the instanton density is parametrically suppressed
 at large $N$ in deconfined region\cite{Parnachev:2008fy},
     \beq
  \label{gamma_N}
 V_{\rm inst}(\theta)\sim e^{-\gamma N} \cos\theta,~~~~ \gamma=\Bigl[\frac{11}{3}
 \ln \left(\frac{\pi T}{\Lambda_{QCD}}\right)-1.86\Bigr].
  \eeq
    It is assumed that  the  higher order   corrections in the instanton background may change the numerical 
   coefficients in $\gamma (T)$.  However,  the general structure of eq. (\ref{gamma_N}) is expected to hold. 
   The critical temperature is determined by condition $\gamma =0$
where  exponentially small expansion parameter  $ \exp{(-\gamma N)}$ suddenly blows up
and becomes exponentially large $ \sim \exp{N}$.
Numerically,
  it happens at 
   \beq
  \label{T_c_N}
  \gamma=\Bigl[\frac{11}{3}  \ln \left(\frac{\pi T_c}{\Lambda_{QCD}}\right)-1.86\Bigr]=0
 ~~~  \Rightarrow ~~~T_c (N=\infty)\simeq 0.53 \Lambda_{QCD},
  \eeq
  where $ \Lambda_{QCD} $ is defined in the Pauli -Villars scheme. 
Our computations are carried out in the regime where the instanton 
density $\sim \exp(-\gamma N) $ is parametrically suppressed  at  any small but finite $\gamma(T)=\epsilon> 0$ when  $N=\infty$.  We identify the regime with $\gamma>0$ with the ``molecular phase" in 4d Euclidean space when the instantons are well defined objects.
From eq. (\ref{gamma_N}) one can obtain the following expression for 
instanton density in vicinity of $T>T_c$, 
   \beq
   \label{T}
   V_{\rm inst}(\theta) \sim \cos\theta \cdot e^{-\alpha N\left(\frac{T-T_c}{T_c}\right)}, ~~~~ 1\gg \left(\frac{T-T_c}{T_c}\right)\gg 1/N.
   \eeq
where $\alpha=   \frac{11}{3}$ and $  T_c (N=\infty)\simeq 0.53 \Lambda_{QCD} $ are estimated at one loop level.
 Such a behavior does  imply that the dilute gas approximation is justified even in close vicinity of $T_c$ as long as $\frac{T-T_c}{T_c}\gg \frac{1}{N}$.     
   Therefore, the $\theta$ dependence, which is sensitive to the 
 topological fluctuations is determined by (\ref{T})  all the way down to the temperatures very 
    close to the phase transition point from above, $T=T_c+ O(1/N)$. 
     The topological susceptibility  
  vanishes
 $\sim   \cos\theta e^{-\gamma N}\rightarrow 0$ for $T> T_c$ in deconfined phase.
  
  The instanton expansion breaks down   at $T<T_c$. In this regime instantons cease to exist, as they dissociate into the instanton quarks, see some references mentioned    in the  Introduction. 
 The instanton quarks with fractional topological charges $1/N$ emerge as  the dominant pseudoparticles in confined phase. 
  This sharp reconstruction  leads to the drastic changes in $\theta$ behaviour. Indeed, the topological susceptibility is order of $\cos(\theta/N)$ in confined phase, while it is of order      $ \cos\theta e^{-\gamma N}$   in deconfined phase. We identify the confined phase with the ``$N$ component plasma phase" in 4d Euclidean space, while deconfined phase  with ``molecular phase" if one uses the  BKT terminology.  
  The strongly coupled regime at $T<T_c$ obviously can not be studied using the  semiclassical approximation. The key observation of  refs.\cite{Parnachev:2008fy,Zhitnitsky:2008ha}    is that we can approach  the vicinities  of the critical point   from above where the semiclassical  expansion remains to be   justified as instanton density is still parametrically small in large $N$ limit as  eq.(\ref{T}) suggests.    This entire picture, in principle, can be theoretically tested  for any finite $N$ using the  weakly coupled ``deformed QCD"  as a toy model  \cite{Shifman:2008ja,Unsal:2008ch}. 
 Such an analysis in principle   allows to study  many deep questions as the $\theta$ dependence, topological susceptibility, and   other related problems   on both sides of the phase transition line,  see  relevant for present work    references  
 \cite{Poppitz:2009uq,Poppitz:2009tw,Thomas:2011ee, Unsal:2012zj,Poppitz:2012nz,Anber:2013sga,Poppitz:2013zqa}.

 Once $T_c$ is determined    one can compute a  finite  segment  of the phase transition line $T_c(\mu)$  for    relatively small $\mu \ll T_c$.  The result    can be presented  as follows~\cite{Parnachev:2008fy} 
\beq
\label{mu}
T_c(\mu)=T_c(\mu=0)\Bigl[1- \frac{3N_f\mu^2}{4N \pi^2 T_c^2(\mu=0)} \Bigr], ~~ \mu\ll \pi T_c, ~~~ N_f\ll N.
\eeq
   This is a solid prediction of the entire framework in large $N$ limit when many  numerical uncertainties 
   \footnote{\label{uncertainties}such as the limitations to the leading loop approximation, uncertainty of the numerical value of $T_c$ at $N=\infty$ in terms of  $\Lambda_{QCD}$ which itself is 
   expressed in terms of the Pauli -Villars scheme, etc}
   are hidden in $T_c(\mu=0)$. It is amazing that     eq. (\ref{mu}) is  in   excellent numerical agreement with lattice data  even for $N=3$ and $N_f=2$ though  eq. (\ref{mu}) 
 was  derived for $N=\infty$, see \cite{Parnachev:2008fy} for the details and references on the original lattice results.  
 
   The same guiding principle can be applied for studying the dense matter as well \cite{Zhitnitsky:2008ha}. To be more specific, we identify    the point where instanton 
 expansion breaks down   with the point $\mu_c$ where the phase transition happens. 
  In the presence of the massless chiral fermions the $\theta$ dependence  obviously  goes away in QCD in both phases: confined as well as deconfined. This  $\theta$- independence of the system in the presence of quarks    reflects  a simple  fact that one can redefine
the fermi fields in the chiral limit such that $\theta$ parameter can be rotated away from the partition function.     
Our conjecture  on sharp changes of the system  in the presence of massless fermions applies to 
the instanton induced potential  itself  which still assumes the same form $\exp{[-\gamma N]}$ similar to the analysis of the pure gauge sysytem (\ref{gamma_N}).
   In our estimates below 
 we assume that the colour superconducting phase is realized in deconfined phase
for all finite $N$.  A precise magnitude of the diquark condensate
 is not  essential  for our calculations as it effects 
 only sub-leading  terms $\sim 1/N$ which will be consistently ignored in what follows, see item {\bf ``c"} below.
 
 The corresponding  estimates for the critical chemical potential $\mu_c$ can be represented as follows \cite{Zhitnitsky:2008ha}:
     \beq
  \label{mu-final}
 \mu_c (N=\infty)\simeq 2.6\cdot \sqrt{\frac{N }{N_f}}\cdot T_c (N=\infty, \mu=0) , ~~~~ N_f\ll N, 
  \eeq
  where we  express the critical value $ \mu_c (N=\infty)$  in terms of the critical temperature (\ref{T_c_N}) to minimize 
  many numerical uncertainties, see footnote \ref{uncertainties}. One should comment that  $\mu_c (N)$ is very large in large $N$ limit and scales as $ \mu_c (N)\sim \sqrt{N}$, in contrast with $T_c\sim1 $. 
  The nature    of this behaviour  can be explained by  the observation that a large number of gluons $\sim N^2$ can get excited at  $T\sim 1$ 
  while  only a relatively small number of quarks in fundamental representation $\sim N$ 
  can get excited at $\mu\sim 1$.   Therefore,   a very large chemical potential $\mu_c^2\sim  {N}$ is required for quarks  to  play a similar role the  gluons  play  at  $T_c\sim 1$. This argument holds  as long as $N_f \ll N$, and all quarks belong to the fundamental representation of $SU(N)$. 
  
 One can also derive the  expression   which is   analogous to eq. (\ref{T}) and which is valid in the vicinity of $\mu_c$ 
 \beq
   \label{mu1}
  V_{\rm inst}  \sim    e^{-\alpha N \left(\frac{\mu-\mu_c}{\mu_c}\right)}, ~~~~ \frac{1}{N}\ll\left(\frac{\mu-\mu_c}{\mu_c}.\right)\ll 1,
   \eeq
Numerically,  the coefficient  $\alpha$ is equal to $11/3$ in the leading order,  similar to expression   (\ref{T}).  Such a behaviour (\ref{mu1}) does  imply that the dilute gas approximation is justified even in close vicinity of $\mu_c$ as long as $ {(\mu-\mu_c)}/{\mu_c}\gg {N}^{-1}$.    In this case the diluteness parameter $\sim   V_{\rm inst} $  remains parametrically small. We can not rule out, of course, the possibility that the  perturbative 
   corrections may change our numerical estimate for $\mu_c$ as well as for $\alpha$. 
However, we   expect that the  qualitative picture of the phase transition advocated by  this 
framework  remains unaffected  as   the corresponding    perturbative  corrections should  be computed, according to our computation scheme,  in deconfined phase where the  instanton  density remains  parametrically small.

One can also study the behaviour of the critical chemical potential $ \mu_c(T) $ as a function of temperature $T$ when it slightly deviates from zero. One arrives 
 to the following expression for a finite segment of the phase transition line  $ \mu_c(T) $.    The result can be presented as follows~\cite{Zhitnitsky:2008ha}:
\beq
\label{T1}
\mu_c(T)=\mu_c(T=0)\Bigl[1- \frac{N \pi^2T^2}{3N_f \mu_c^2(T=0)} \Bigr], ~~~~
 \sqrt{N}T\ll \mu_c, ~~~~ N_f\ll N.
\eeq
 It is expected that the phase transition line (\ref{T1}) computed at small $T$ and large $\mu_c$ continuously  connects  to 
  the phase transition line (\ref{mu}) computed at small $\mu$ and large $T_c$.

 We conclude this short overview with the following important comments. 
There are three  basic reasons for a generic structure
(\ref{gamma_N})  to emerge:\\
{\bf a.} The presence of the exponentially large ``$T-$ independent"  and ``$\mu -$ independent" contributions
( e.g. ~$e^{+1.86 N}$ in eq. (\ref{gamma_N})). This term 
   basically describes  the entropy of the configurations and enters the exponent with the positive sign. It is due to a number of contributions such as a 
 number of embedding $SU(2)$ into $SU(N)$ etc;\\
 {\bf b.} The  presence of the ``$T-$ dependent" and ``$\mu-$ dependent" contributions to  the instanton induced potential $V_{\rm inst}  $ which come  from  $\int n(\rho) d\rho$ integration.   In case of  $T\neq $ and $\mu=0$ it   is proportional to
  \beq 
  \label{rho}
     \left(\frac{\Lambda_{QCD}}{\pi T}\right)^{\frac{11}{3}N}=\exp\Bigl[-\frac{11}{3}N
\cdot \ln \left(\frac{\pi T}{\Lambda_{QCD}}\right)\Bigr].
 \eeq
 This term enters the exponent with the negative sign because  the corresponding contribution must be strongly suppressed
  for large $T$ and $\mu$; \\
 {\bf c.} The fermion- related contributions such as a  chiral condensate, diquark condensate  or non-vanishing mass term    enter the instanton density as follows $ \sim\la\bar{\psi}\psi\ra^{N_f}\sim e^{N\cdot \left(\kappa\ln  |\la\bar{\psi}\psi\ra|\right)} $. For $\kappa\equiv\frac{N_f}{N}\rightarrow 0$ this term  obviously leads to a sub leading effects $1/N$ in comparison
  with two  main terms in the exponent (\ref{gamma_N}). Therefore, such terms can be neglected as they do not change any estimates
  at $N=\infty$. It is  in accordance with the general arguments suggesting that the fundamental fermions can not change the dynamics of the relevant gluon configurations  as long as $N_f\ll N$. 
   
The crucial element in this  analysis   is that both leading contributions (items {\bf a} and {\bf b} above) have exponential $\exp(N)$ dependence,
and therefore at $N\rightarrow \infty$ for $T>T_c$ the instanton gas is dilute with density
$\exp{(-\gamma N)}, ~\gamma>0$ which ensures  the behaviour  (\ref{T}).  For  $T<T_c$ the density is exponentially large 
$\sim\exp(+N)$ which implies that 
the instanton expansion breaks down. We expect that the instantons will dissociate to the constituents, the instanton quarks,  
which become  the dominant pseudo-particles at $T<T_c$. This happens exactly at the point of the phase transition when the complete reconstruction of relevant pseudo-particles occurs, in close analogy with BKT-like transition.   The key observation here is that  our predictions are {\it not very  sensitive to  a  precise mechanism} of this reconstruction for large $N$.  In other words,  one can obtain  a number of solid relations which follow from  this framework, such as (\ref{mu}), (\ref{mu-final}), (\ref{T1}),   without a detail  knowledge of the dynamics describing  the  instanton's dissociation in the large $N$ limit.

 \section{Deconfinement   phase in QCD at large $N\gg 1$ and $N_f\sim N$.} \label{conformal}
   Our goal here is to generalize the ideas reviewed in section \ref{review} to include into the system a large number of fermions $N_f\sim N$.  The corresponding description can be conveniently  expressed in terms of parameter $\kappa\sim 1$. As we mentioned in the Introduction 
we wish to study the correlation function  $\la {\rm det }~ \bar{\psi}_L^f \psi_R^f (x_i)\ra$ 
as a function of parameter $\kappa$ and distances $(x_i-x_j)$.  Our main criteria  to pinpoint the phase transition line remains the same as formulated above: we study the instanton induced potential as a function of $\kappa$. At small $\kappa\ll 1$ the integral over $\rho$ diverges in the infrared,  which corresponds to the confined phase when the dominant pseudo-particles are the fractional  instanton quarks. This corresponds to the strongly coupled regime, the ``plasma phase" in BKT terminology. This regime by obvious reasons  can not be studied by semiclassical methods used in the present work.  As we show below for  $\kappa > 1$  the integral over $\rho$ starts to  converge at large $\rho$. It is necessary, but obviously not  a sufficient, condition for the transition to the ``molecular phase" in BKT terminology. The sufficient condition for the ``molecular phase" to set in,  according to our conjecture, is   a sudden $\exp (-N)$ suppression of the instanton density in the deconfined phase similar to eqs (\ref{T}), (\ref{mu1}). This is exactly the point when   the constituents   coalesce  by forming a finite size instanton. In other words, 
the constituents merge and get bounded   into a single instanton organizing the  ``molecular phase" in BKT terminology. We identify this point with the critical $\kappa_c$ which is the minimal value of $\kappa$ where the ``conformal window" starts, see the left point  of the thick black segment   plotted  on Fig. \ref{phase}. We shall also study the variation of this critical value  $\kappa_c (\mu, T)$ with small variations of the chemical potential $\mu$ and temperature $T$. 
\subsection{Computations and   technical details}
We start the implementation  of this program by analyzing  the convergence properties  of the $d \rho$ integral at large $\rho$.  We use the standard formula
for the instanton density at one -loop order \cite{tHooft,shuryak_rev}
\begin{eqnarray}
\label{instanton}
 n(\rho)= C_N(\beta(\rho))^{2N} \rho^{-5}
 \exp[-\beta(\rho)]  
 \times  \exp[-(N_f \mu^2 + \frac13 (2N+N_f) \pi^2
 T^2)\rho^2], 
\end{eqnarray}
where
\begin{eqnarray}
\label{beta}
  C_N &=& \frac{0.466 e^{-1.679N} 1.34^{N_f}}{(N-1)!(N-2)! },~~~~
\beta(\rho)=-b \ln(\rho\Lambda_{QCD}), ~~~~  b=\frac{11}3 N-\frac23 N_f
\end{eqnarray}
This formula  contains, of course, the standard instanton classical action $\exp(-8\pi^2/ g^2(\rho))
\sim  \exp[-\beta(\rho)]  $
which however is hidden as it is   expressed in terms of $\Lambda_{QCD}$
rather than in terms of coupling constant $g^2(\rho)$.  We inserted the   chemical potential $\mu=\mu_B/N$ and temperature $T$ into this  expression for future consideration when we shall study the critical value $\kappa_{c} (T, \mu)$ as a function of $T, \mu$ at small values of these external parameters.

We start with $T=\mu=0$. In the chiral limit 
the integral over $\rho$ is reduced to the following expression
\beq
\label{det}
\la {\rm det }~ \bar{\psi}_L^f \psi_R^f  (x_i)\ra= \int d\rho n(\rho)d^4x\prod_i^{N_f}\frac{2\rho^3}{\pi^2
\left[(x-x_i)^2+\rho^2\right]^3},
\eeq
 where we keep only zero modes in the chiral limit, assuming that in the dilute gas approximation (which will be  justified for sufficiently large $\kappa$ as we shall argue below) all other mode contributions is suppressed by factor $m_q\rightarrow 0$. The integration over $d^4x$ corresponds to the integration over the instanton center at point $x$. 
 
 In order to study the behaviour of the integral at large $\rho$ we take $x_i=x_j$ and integrate over $d^4x$.    One arrives to the following expression
 \begin{eqnarray}
 \label{det_1}
\la {\rm det }~ \bar{\psi}_L^f \psi_R^f\ra  =\frac{ \pi^2}
{(3N_f-1)(3N_f-2)} \int\!d\rho\, n(\rho) 
\rho^4\cdot \biggl(\frac{2}{\pi^2 \rho^3} \biggr)^{N_f} , 
  \end{eqnarray}
where $n(\rho)$ is defined as before by eq.(\ref{instanton}). The combination  
 $ \int\!d\rho\, n(\rho) \rho^4$ is dimensionless while 
 the dimension of the operator $\la {\rm det }~
  \bar{\psi}_L^f \psi_R^f\ra\sim \la \rho\ra^{-3N_f}\sim$
  (MeV)$^{3N_f}$ as it should. The next step is to evaluate the integral at large $N$ 
  (and keeping $\kappa=N_f/N$  fixed) by 
  using the standard Stirling formula 
    \beq
  \label{stirling}
  \Gamma (N+1)=\sqrt{2\pi N}N^N e^{-N}\left(1+\frac{1}{12N} +O(\frac{1}{N^2})\right)
  \eeq
  One   follows the same procedure as before  \cite{Parnachev:2008fy,Zhitnitsky:2008ha} to evaluate integral $\int d\rho$ and   take the limit $N\rightarrow\infty$  at the end of computation. In the present case $\kappa=\frac{N_f}{N}\sim 1$ one should keep few additional numerical factors such as 
  $1.34^{N_f}$ from  (\ref{instanton}) and take into account changes in $b$ due to $N_f$ in eq. (\ref{instanton}) which have been previously \cite{Parnachev:2008fy,Zhitnitsky:2008ha} ignored.
  However, the most drastic change occurs due to large power of  $\rho^{-3N_f}$ in eq. (\ref{det_1}).
  Collecting all leading order contributions we arrive to the following expression for the  expectation value $\la {\rm det }~ \bar{\psi}_L^f \psi_R^f\ra$ in the large $N$ limit at the one-loop level:
   \begin{eqnarray}
 \label{estimate}
\la {\rm det } ~\bar{\psi}_L^f \psi_R^f\ra  \sim \Lambda_{QCD}^{N(\frac{11}{3} -\frac{2}{3}\kappa)}
\int_{1/M}^{\infty} \frac{ d\rho}{\rho}\rho^{\frac{11}{3}N\cdot (1-\kappa)} \cdot \exp{\left[N\left(-1.679+2+2\ln(\frac{11}{3}-\frac{2}{3}\kappa)+1.34 \kappa-\kappa\ln(\frac{\pi^2}{2})\right)\right]}.
  \end{eqnarray}
  In our estimate (\ref{estimate}) we neglected all $ \ln^n(\rho\Lambda_{QCD})$ contributions which enter $\int d\rho$ integral.  First,  the corresponding logarithmic  contributions do not change the convergent properties of the integral, which is the main element   of our studies in the present work. Second, the corresponding numerical uncertainty  of the integral due to  $ \ln^n(\rho\Lambda_{QCD})$ contributions  can be effectively incorporated into  the parameter $M$, which itself is not numerically known as a result of an additional non-perturbative UV divergences\footnote{We should remark here that a consistent procedure to collect all $ \ln^n(\rho\Lambda_{QCD})$ factors   requires the two-loop computations in the exponent in eq. (\ref{estimate}) as $\exp[ \ln\ln(\rho\Lambda_{QCD})] = \ln(\rho\Lambda_{QCD})$, which is beyond the scope of the present work. Furthermore, the corresponding numerical value of $\Lambda_{QCD}$ should be expressed  at  the two-loop level   for the consistency of this procedure.  The corresponding numerical uncertainties can be incorporated in terms of the ratio $\ln(M/\Lambda)$ as we discuss below. We already   mentioned in section \ref{review} that the higher loop corrections in the instanton background obviously modify all numerical estimates within this framework.  However, we do not expect that these higher loop corrections can drastically modify the qualitative picture of the phase transition advocated in this work.}, see below few comments on physical meaning of parameter $M$.  All other numerical factors in the brackets in the exponent in eq.(\ref{estimate}) can be easily traced from the original expressions  (\ref{instanton}), (\ref{beta}), (\ref{det}). \exclude{Indeed, factors $-1.679N$ and $1.34 \kappa N$   follow from the original expression for the instanton density (\ref{beta}), while the factor $2N\ln(\frac{11}{3}-\frac{2}{3}\kappa)$ enters the instanton density $b^{2N}$ in eq. (\ref{instanton}). Finally, the numerical factor $-N\kappa\ln(\frac{\pi^2}{2})$ appears as a result of normalization of the zero modes (\ref{det}), while  
  contribution  $2N$ enters  eq.(\ref{estimate})  as a leftover  of strong cancellations of many $N!$ factors  from  the instanton density  (\ref{beta}).  }

  \exclude{   
  such that the most relevant part of integrand   takes the form
  \begin{eqnarray}
 \label{estimate_0}
\la {\rm det } ~\bar{\psi}_L^f \psi_R^f\ra  \sim \Lambda_{QCD}^{N(\frac{11}{3} -\frac{2}{3}\kappa)}
\cdot e^{N(5.65-2.63 \kappa)}
\int_{1/M}^{\infty} \frac{ d\rho}{\rho}\rho^{\frac{11}{3}N\cdot (1-\kappa)},
  \end{eqnarray}
  where for simple estimates we neglected all $(\ln\rho)^n$ contributions as well as all powers $N^n$
  in   the integrand.    The dependence on $\kappa$ in the exponent in (\ref{estimate}) is,  in fact,  quite complicated function of $\kappa$ as one can see from the expression for the $\beta$ function (\ref{beta}). It includes, in particular,  $\ln b$ which depends on $\kappa$. However, for our purposes in the relevant for us region of $\kappa$ it is sufficient to approximate this behaviour   by a simple linear function of $\kappa$ in the exponent as shown in eq. (\ref{estimate}). 
 }
 
 Few comments are in order.
 First of all, the integral (\ref{estimate}) is divergent in the ultraviolet (UV) for $\kappa \geq 1$  at $\rho\rightarrow 0$. This  
 UV  divergence has a non-perturbative origin, which was noticed for the first time in \cite{Shuryak:1987an}, see also \cite{Velkovsky:1997fe} with related discussions. This UV divergence 
 should be  contrasted with conventional perturbative UV divergences in quantum field theory (QFT)
 which   normally emerge  when the QFT operators are defined at coinciding points.  
 The UV divergence (\ref{estimate})  has no relevance for our studies as we are interested in the convergence properties  of this integral in the IR at $\rho\rightarrow \infty$, rather than in UV,  as explained in the Introduction and section \ref{review}.   Nevertheless, we have to deal with this non-perturbative UV divergence as its finite portion explicitly enters the expression for the vacuum expectation value $\la {\rm det } ~\bar{\psi}_L^f \psi_R^f\ra$.  We use  conventional for QFT renormalization procedure by subtracting the corresponding UV divergence and introducing a new dimensional parameter, the point of normalization $M$ where  $\la {\rm det } ~\bar{\psi}_L^f \psi_R^f\ra_M$ is defined.  Essentially, we fix the magnitude of the expectation value of the operator (\ref{estimate}) in terms of parameter $M$. We further elaborate on the physical meaning of this  parameter $M$ later in the text. 
    
  Our next comment is as follows:  when $\kappa>1$ the integral is convergent in the IR for large $\rho$. The convergence of the integral at large $\rho$ is necessary 
  but not sufficient condition for the instantons to be in the dilute gas regime. Similarly, the integral over $\int d\rho$ is convergent for arbitrary small temperature $T$, but it does not imply, of course,  that the instanton density (\ref{gamma_N}) is small.
  Regime of exponentially small instanton density  is achieved when $ T$ in eq. (\ref{gamma_N}) is sufficiently large 
  and $\gamma(T)$ flips the sign. Equation  (\ref{estimate}) which is intended   to study the conformal window exhibits a   similar behaviour  for sufficiently large $\kappa$, as we shall argue below. 
    
  Now we return to eq. (\ref{estimate}) to elaborate on   the physical meaning of the cutoff parameter $M$.  For sufficiently large $\kappa$ 
  the expression (\ref{estimate})  becomes exponentially small. We associate  this   regime (which is characterized by an  exponentially small instanton density) with the   deconfined phase similar to our studies of the deconfined phase transition at  high temperture $T>T_c$  \cite{Parnachev:2008fy}  and high chemical potential $\mu > \mu_c$ \cite{Zhitnitsky:2008ha}. The chiral symmetry is also expected to be restored in this phase  as the chiral condensate 
 normally  forms  as a result of strong interactions between the instantons, which obviously can not be the case as the instanton density is  exponentially suppressed in this regime. This conclusion is   supported by  the  computations in the instanton liquid model
  which also suggest that the chiral condensate vanishes when sufficient number of flavours are present in the system~\cite{Velkovsky:1997fe}. According to Fig. \ref{phase} we identify this deconfined phase with the conformal window. In conformal field theory  $\Lambda_{QCD}$ entering  the expression (\ref{estimate}) is a fictitious scale  as there is no dynamically generated scale in the system in this phase. Scale $M$ was  introduced as point of normalization for the operator  eq. (\ref{estimate}) to remove the non-perturbative UV divergences. It can be also   thought as an effective  scale (in the conformal window phase)  at which the running constant is saturated to its infrared value. Indeed, the  $\rho$ dependence  in eq. (\ref{estimate}) is essentially originated from  $g(\rho)$ dependence entering the instanton density (\ref{instanton}, \ref{beta}).  Therefore, saturation of the integral (\ref{estimate}) at $\rho\sim M^{-1}$ can be interpreted as  saturation of the coupling constant at this scale, i.e. 
  $g(\rho)\simeq g(M)\simeq g_{IR}$, where $g_{IR}$ is the  coupling constant at the IR fixed point.    The  $\Lambda_{QCD}$ in this system is not an independent parameter, but proportional to $M$,  i.e. $  \Lambda\sim M $, where we dropped subscript $_{``QCD"}$  in $\Lambda$ to emphasize that there is no any dynamically generated scales in this phase.  Furthermore, as we discuss below, 
  the correlation function  $\la {\rm det } ~\bar{\psi}_L^f \psi_R^f (x_i)\ra$ exhibits the power like decay $(x_i-x_j)^{-\gamma_{\rm det }}$ with nontrivial 
   anomalous dimension $\gamma_{\rm det }$. It implies that the distances $(x_i-x_j)$ are measured in units of the same scale $M\sim \Lambda$.

  To get some feeling about numerical values of relevant parameters when the conformal window may emerge in this framework,   it is convenient to  represent the integral (\ref{estimate})  in the following form
  \begin{eqnarray}
 \label{estimate1}
\la {\rm det } ~\bar{\psi}_L^f \psi_R^f\ra  \sim \Lambda^{3N\kappa} \cdot \exp{\left[-N\gamma(\kappa)\right]}, ~~
 \gamma (\kappa)= \left[ \frac{11}{3}\ln \left(\frac{M}{\Lambda}\right)-5.65\right]-\kappa\left[ \frac{11}{3}\ln \left(\frac{M}{\Lambda}\right)-2.63\right] , ~~ \kappa>1, 
  \end{eqnarray}
  where we expanded the $\ln(\frac{11}{3}-\frac{2}{3}\kappa)$ function in vicinity of $\kappa\simeq 4$ to simplify things and to display 
  the most important features of the $ \gamma (\kappa)$ function where the phase transition is expected to occur according to the lattice studies, see review paper \cite{Pallante:2009hu}. 
In this simplified form eq. (\ref{estimate1})  assumes the structure  similar   to eq. (\ref{gamma_N}). One can follow exactly the same logic  as before  to search for a condition when  the instanton density   suddenly becomes exponentially suppressed $\sim \exp(-N)$ which is identified with deconfined phase transition.
  However, there is a crucial difference between present studies represented by eq.(\ref{estimate1}) and  previously analyzed  cases $T>T_c$ and $\mu>\mu_c$ reviewed in  section \ref{review}.  The point is that in our previous studies we had a single unknown parameter such as $T_c$ which was expressed in terms of $\Lambda_{QCD}$ defined in the Pauli -Villars scheme (\ref{T_c_N}). In present case we have two unknown parameters: $\kappa_c$ and $M$. Therefore, a single condition that $\gamma(\kappa)$ in eq. (\ref{estimate1}) flips the sign at the critical value $\kappa_c$ 
  is not sufficient to estimate $\kappa_c$. If, instead, we 
    adopt $\kappa_c\simeq 4$ as suggested by the lattice simulations at $N=3$, see review paper \cite{Pallante:2009hu},
   we infer that ${M}/{\Lambda}$ should be ${M}/{\Lambda} \simeq 1.55$ in eq. (\ref{estimate1}) for $\gamma(\kappa)$  to flip the sign at $\kappa_c\simeq 4$.  Of course, it can  not be considered as a prediction of our framework. Rather, it should be considered as  the consistency check
   that equation $\gamma(\kappa_c)=0$ has a solution       with very reasonable physical parameters.  While eq. (\ref{estimate1}), in contrast with previously considered   cases,   has not produced an unambiguous  prediction for $\kappa_c$ 
   we shall see in a moment that there are in fact few solid consequences  of this framework, which can be in principle tested in the lattice simulations. 
      
   One can easy understand the nature of the  additional uncertainty (which was not   present in  previous cases)    entering (\ref{estimate1})  in the form of dimensionless parameter ${M}/{\Lambda}$. Indeed, in our previous studies any uncertainties related to a magnitude of  diquark  condensate, or specific properties of quarks, etc, could only  produce a parametrically small  effect $\sim N_f/N\rightarrow 0$ for finite $N_f$ and large $N\gg 1$, see item {\bf c)} 
   in section \ref{review}. In contrast, in present case when we study the vacuum expectation value of the operator $\la {\rm det } ~\bar{\psi}_L^f \psi_R^f\ra$ with  dimensionality 
     of order of $N$ the uncertainty  enters (\ref{estimate1}) at the leading order  $\sim N$.   Precisely this uncertainty prevents us from making a solid prediction for $\kappa_c$ in large $N$ limit within this framework.
   Nevertheless,  there are few other firm  consequences   of this approach to be    discussed below. 
   
   \subsection{The basic consequences of the framework}
   While we can not  predict the position of the critical value  $\kappa_c$, we can study the behaviour of the system in close vicinity of the critical point in large $N$ limit. The corresponding expression follows from (\ref{estimate1}) and reads
    \begin{eqnarray}
    \label{kappa}
   \la {\rm det } ~\bar{\psi}_L^f \psi_R^f\ra  \sim \Lambda^{3N\kappa} \cdot 
   \exp{\left[-3\cdot N\cdot \left(\frac{\kappa-\kappa_c}{\kappa_c-1}\right)\right]}, ~~~~~~~~ 
   \frac{1}{N}\ll \left( {\kappa-\kappa_c} \right)\ll 1.
    \end{eqnarray}
This formula is very similar in all respects to previously discussed cases (\ref{T}), (\ref{mu1}) when one approaches the critical point from the  deconfined side of the phase transition line. The behaviour (\ref{kappa}) implies  that the dilute gas approximation is justified even in close vicinity of $\kappa_c$ as long as $  \left( {\kappa-\kappa_c} \right)\gg 1/N $.   The  instanton 
density $\sim \exp(-N) $ is parametrically suppressed  in this region for   any small but finite $\left( {\kappa-\kappa_c} \right)=\epsilon> 0$ for large   $N$.  We identify the regime   with the ``molecular phase" in BKT terminology when the instantons are well defined objects with finite sizes. 
This deconfined regime is identified according to Fig. \ref{phase} with the conformal window.  When we cross the phase transition line the instanton expansion breaks down   at $\kappa<\kappa_c$. In this regime instantons cease to exist, as they dissociate into their constituents, the instanton quarks,  which become the dominant pseudo-partcile, see some references mentioned    in the  Introduction. This regime corresponds to the confined phase where the semiclassical approximation can not be trusted, and we do not even  attempt to make any computation   in this regime at $\kappa<\kappa_c$.

The behaviour (\ref{kappa}) is very generic feature of this framework
which explicitly shows that the transition from one phase to another is very sharp at large $N$. This behaviour becomes even more pronounced with increasing $N$, in close analogy with the similar analysis   (\ref{T}) when the temperature $T$ varies.  In the case with temperature's variations    the corresponding lattice simulations \cite{Alles:1996nm} -\cite{ Vicari:2008jw} strongly support such  drastic changes of the topological fluctuations.    We expect exactly the same type of behaviour with variation of  $\kappa$  as the physics behind of eq. (\ref{T}) and eq.(\ref{kappa}) is exactly the same: it is complete reconstruction of the dominant pseudo-particles   when $\kappa$ varies  from deconfined regime with  $\kappa=(\kappa_c+\epsilon)$ to confined phase with $\kappa=(\kappa_c-\epsilon)$.

Now we turn  to analysis  of the critical behaviour  $\kappa_{c} (T, \mu)$ when 
   $T$ and  $\mu$ slightly deviate from zero.  We still remain  in the conformal region where the instanton based computations are  under complete theoretical control. In this case a simple one-loop insertion
$   \exp[-(N_f \mu^2 + \frac13 (2N+N_f) \pi^2
 T^2)\rho^2] $ in eq. (\ref{instanton}) is justified. We consider this term as a perturbation  to the instanton density which is still exponentially suppressed. 
 This term obviously makes  the instanton density even smaller. Therefore the transition  between confined and deconfined phases happens
 at a smaller critical values of $\kappa_c(T, \mu)$ in comparison with  $\kappa_c(T= \mu=0)$.
 To  quantify  this argument we notice that  the  integral (\ref{estimate}) is saturated (for small $T$ and $\mu$) by  $\rho\sim M^{-1}$  as before.
Therefore,  we arrive to the following numerical estimate for this variation of  $\kappa_c(T, \mu)$ in comparison with its 
$\kappa_c(T= \mu=0)$ value:
 \beq
 \label{T_mu}
 \Bigl[\kappa_c(T, \mu) -\kappa_c(T= \mu=0)\Bigr] \simeq -0.3 \cdot (\kappa_c-1)\cdot
\frac{ \left(\kappa_c \mu^2 + \frac13 (2+\kappa_c) \pi^2T^2\right)}{M^2}, ~~~ T, \mu \ll M.
 \eeq
Equation (\ref{T_mu}) is a direct consequence of the entire framework. It is  testable, in principle,   on the lattice.
Essentially, eq (\ref{T_mu})    predicts how the conformal window (shown on Fig. \ref{phase}) expands at small, but non-vanishing $T, \mu \ll M$.  We note here that eq (\ref{T_mu})  is a close  analog of  the previously  discussed  eqs. (\ref{mu}) and (\ref{T1}).  As we already mentioned the corresponding expression (\ref{mu}) is in very good agreement with available lattice results. 
 
   Now we want to demonstrate that the correlation function (\ref{det}) exhibits  a power like decay  at large $(x_i-x_j)$  with a specific critical exponent  $\gamma_{\rm det}$. Such a behaviour is    the  distinct sign of the  conformal phase. Indeed, the integral 
    (\ref{det})  is convergent in the IR at large $\rho$,  and in fact is numerically exponentially small $\exp(-N)$ for large $N$ in deconfined phase at  $\kappa>\kappa_c$ as we argued above.
    Furthermore, it is perfectly convergent in the UV at $\rho\rightarrow 0$   for  non-coinciding points $x_i\neq x_j$. We are interested in the behaviour of this correlation function\footnote{this function obviously has a pure non-perturbative origin, and can not be generated in the perturbation theory} when all distances are the same order of magnitude $\sim R$ and large, i.e. $(x_i-x_j) \sim R\gg M^{-1}$.  In this case the integral (\ref{det}) obviously exhibits the algebraic  decay as the integral  is saturated by large $\rho\sim R$. The critical behaviour determined by  $\gamma_{\rm det}$  of the correlation function $\la\bar{\psi}_L^f \psi_R^f (x_i)\ra $ can be inferred    from a simple dimensional argument. It is given by 
 \beq
  \label{decay}
\la {\rm det } ~\bar{\psi}_L^f \psi_R^f (x_i)\ra  \sim {\Lambda^{N(\frac{11}{3} -\frac{2}{3}\kappa)}\cdot {\cal{P}} \left[(x_i-x_j), \gamma_{\rm det} \right]}\sim  \Lambda^{N(\frac{11}{3} -\frac{2}{3}\kappa)}\cdot \frac{1}{R^{\gamma_{\rm det}}}, ~~~ \gamma_{\rm det}=\frac{11}{3}N\cdot (\kappa-1), ~~~ \kappa\geq \kappa_c, 
  \eeq
  where ${\cal{P}} [ (x_i-x_j), \gamma_{\rm det}] $ in general  is a complicated  algebraic function of variables  $(x_i-x_j)$. Indeed, as one can see from the original expression (\ref{det}) it must depend only on relative distances $(x_i-x_j)$ as a result of translation invariance.
Furthermore, ${\cal{P}} [ (x_i-x_j), \gamma_{\rm det}] $ must be symmetric under exchange 
  of the external points $x_i\leftrightarrow x_j$ as the original expression (\ref{det}) exhibits this  symmetry.  
  
  Few comments are on order. First,  the decay law (\ref{decay}) should be compared with   behaviour of this  correlation function  with  dimension $\gamma_{\rm canonical}=3N_f$ which corresponds to the decay $R^{-3N_f}$ representing the    canonical behaviour of free massless fermions. The scaling  behaviour    (\ref{decay}) always exhibits a slower decay power at  large distances  
  than the canonical decay  as long as the asymptotic freedom holds, i.e.  $(\frac{11}{3}N -\frac{2}{3}N_f)>0$. We should also comment that the $(\ln \rho)^n$ corrections which enter the expression for the instanton density (\ref{instanton},\ref{beta}) and which were consistently ignored in our computations  will modify a simple scaling property (\ref{decay}) by producing $(\ln R)^n$ corrections in front of eq. 
      (\ref{decay}). Such type of corrections are expected, and in fact, are very generic for conformal theories.  
       Finally, in large $N$ limit when $N_f\sim N$ it is instructive to represent the critical exponent  $\gamma_{\rm det}$ in terms of anomalous dimension   per number of flavours, i.e.
  \beq  
  \label{gamma}
  \frac{\gamma_{\rm det}}{N_f}=\frac{11}{3}\frac{ (\kappa-1)}{k}, ~~~ \kappa\geq \kappa_c.
\eeq
   Formula (\ref{gamma}) shows 
   that a properly normalized critical exponent approaches the finite  magnitude  in large $N$ limit. The corresponding limiting value  in this approach is unambiguously determined by $\kappa$.

   We conclude this section with the following  general comment. We advocate a picture  that the confinement- deconfinement 
   phase transition is a result of complete reconstruction of the dominant pseudoparticles in the system.  The confined phase can be treated as the ``plasma phase" in BKT terminology when the instanton's constituents are de-localized pseudoparticles,  while the deconfined phase can be thought as the ``molecular phase" when $N$ coherent instanton's constituents are well-localized and well-organized in a form of a single    instanton with a finite size. As we already  emphasized in the Introduction,  our system is formulated   in terms of pseudo- particles which live in 4d Euclidean space rather  in terms of real statical particles/quasiparticles which live in Minkowski space-time.   Therefore, one should not confuse the thermodynamical terms such as ``phase", ``density" or ``entropy" applied to this 4d statistical ensemble with  the corresponding entities   which are normally applied to a  conventional system  of static objects in 3+1 Minkowski space.    Nevertheless, one can interpret   the exponential suppression $\exp(-N)$ of the ``density" of the pseudo-partciles in the ``molecular phase"    as a result of this re-organization of the dominant pseudoparticles in the system.
   
    This picture of complete reconstruction of the dominant pseudo-particles presented above  
  strongly  resembles the conventional BKT formulation    of the 2d XY model ~\cite{BKT}.  Indeed, 2d XY model  can be described at low temperatures as the system       when the vortex and anti-vortex pairs  make the bound states.
  This phase  corresponds to   the conformal field theory with infinite correlation length. At the same time at high  temperatures the vortices 
  and anti-vortices are not bounded and can be described as a 
   two component  plasma    with a finite correlation length~\cite{BKT}.  This behaviour in conventional BKT system with real physical particles is ``opposite"  to what we observed in    the system of 4d objects when the high temperature phase ($T>T_c$ or $\mu > \mu_c$ or $\kappa> \kappa_c$) corresponds to  the ``molecular phase" of the bounded objects (instantons),  while the low temperature phase ($T<T_c$ or $\mu < \mu_c$ or $\kappa< \kappa_c$) corresponds to the ``plasma phase" of unbounded objects (instanton quarks). This ``opposite" trend is also   quite generic feature of the 4d Euclidean objects. For example, the contribution of the 4d instanton quarks/instantons  to the   topological susceptibility is always opposite in sign in comparison with the corresponding contributions of any physical  propagating degrees of freedom, see e.g.   \cite{Thomas:2011ee} 
   where   computations in a weakly coupled gauge theory can be explicitly performed.

    The conventional BKT picture    also suggests that the correlation length $\zeta$ being expressed in terms of parameter $(\kappa-\kappa_c)$
   should exhibit a very specific behaviour $\zeta\sim  \exp(\frac{1}{\sqrt{\kappa_c-\kappa}})$ when $\kappa$ approached $\kappa_c$ from the plasma phase side, see comment in footnote \ref{BKT}. Now we can explicitly see from eq. (\ref{kappa}) why  the correlation length $\zeta$  at $\kappa<\kappa_c$  can  not  be  studied   in our approach:  the instanton expansion blows up   in this region.   The break down  of the instanton expansion  is formally expressed in terms of  exponentially large density   $\sim \exp(N)$. The corresponding studies in the ``plasma phase" require fundamentally different description  of the problem formulated from the very beginning in terms of the instanton's constituents rather than instantons themselves, see also a related comment in the Conclusion.

   \section{Conclusion and future directions}\label{conclusion}
   In this paper we advocate an idea  that the confinement-deconfinement phase transition in large $N$ limit
      can be formulated  as dissociation of an  instanton into its     constituents, the so-called instanton quarks, see footnote  
 \ref{constituents} with description of some generic properties of the constituents, and also with few historical remarks on the subject. In the ``molecular phase" the constituents are bound into a finite size instanton
 while in  the ``plasma phase"   each instanton dissociates into $N$ delocalized objects, the instanton quarks.   This picture strongly resembles the well -known BKT formulation  of the 2d XY model  
 as mentioned above. 
  
 This idea  was (successfully) applied previously to the confinement-deconfinement phase transition when temperature $T$ and chemical potential $\mu$ vary. The main objective of the present work is to apply the same  ideas and the same guiding principles to study the phase transition at large $N$ when number of flavours $N_f$ (formulated in terms of $\kappa=N_f/N$) varies. We find that for sufficiently large $\kappa$ the phase transition does occur, and we argue that 
 for $\kappa>\kappa_c$ the ``molecular phase" in BKT terminology is in fact a conformal field  theory, similar to the original BKT  
 studies. The main results of this framework are expressed by  eqs. (\ref{kappa}, \ref{T_mu}, \ref{decay}, \ref{gamma}).
 The key observation here is that our predictions in large $N$ limit are not very sensitive to a precise mechanism of dissociation of an  instanton to its constituents. In other words, one can obtain a number of solid relations   (\ref{kappa}, \ref{T_mu}, \ref{decay}, \ref{gamma})  without a detail knowledge of the dynamics describing the instantonÕs dissociation into its constituents, which is indeed a  very
 hard problem of strongly interacting system. We do expect that numerical coefficients in eqs. (\ref{estimate}, \ref{estimate1}, \ref{kappa}, \ref{T_mu}) may change as a result of higher order perturbative corrections in the instanton background. However, we do not expect that entire picture advocated in the present work may experience any drastic changes. 
 
 In fact, the basic framework is supported not only by the original arguments   motivated by  the holographic QCD.  It is also  supported by the lattice simulations and, though implicitly, by   recent computations in the so-called ``deformed QCD", as we already mentioned in the Introduction. As we   emphasized previously, the mechanism  of the phase transition   advocated in the present work can be viewed as  $\grave{\rm a}$ la   BKT 
 phase transition  when   in the ``molecular phase" the constituents are bound into finite size instantons 
 while in  the ``plasma phase"   each instanton dissociates into $N$ delocalized objects, the instanton quarks. Apparently, this picture     is quite universal and looks very much the same independently on a specific property of the parameter which varies: the temperature $T$, the chemical potential $\mu$ or the number of flavours $N_f=\kappa N$. 
 We conclude this work with    some thoughts on possible  future directions related to the present studies: 
 
$\bullet$ The key observation of this work   is that we can approach the phase transition point $\kappa_c $ with sufficiently high accuracy  in large $N$ limit  when the instanton induced potential is exponentially small 
 $\exp (-N)$ and our computations  are justified (\ref{kappa}).   In other words, we are not very sensitive to some hard  problems    related to the dynamical description  of the  dissociation of the instantons to their constituents. This physics is taking place   precisely in the window   which is not accessible
 within our framework:  $(\Delta T)/T\sim 1/N$, see eq. (\ref{T}),   or $\Delta\kappa\sim 1/N$, see eq. (\ref{kappa}). At large $N$ this unaccessible window shrinks to a point. However, in nature    $N=3$, which is not very large. Therefore, the window $(\Delta T)/T$ numerically  could be order of unity, and therefore, this region could be phenomenologically very important.  In other words,  the window  $(\Delta T)/T$ can  be experimentally  studied in   relativistic heavy  ion collisions. In fact,   some  observations and numerical studies  apparently suggest that there are additional magnetic degrees of freedom  which may   play a  role in thermodynamics of the quark gluon plasma slightly above $T_c$, see \cite{Chernodub:2006gu,D'Alessandro:2007su,Chernodub:2009hc,D'Alessandro:2010xg,Bonati:2013bga}  and review \cite{Shuryak:2008eq}. The corresponding discussions are beyond the scope of the present work.   However, we would like to mention here that the instanton constituents carry the magnetic charges, and they should be considered as the excitations  which may contribute into the thermodynamics of the system at $T>T_c$. It would be very exciting if the corresponding magnetic degrees of freedom discussed in \cite{Chernodub:2006gu,D'Alessandro:2007su,Chernodub:2009hc,D'Alessandro:2010xg,Bonati:2013bga,Shuryak:2008eq} can be identified 
with the instanton's constituents, in which case the corresponding magnetic degrees of freedom must condense exactly at the phase transition point $T_c$. At  the temperatures slightly above $T_c$  the constituents can move as they  are not sufficiently strongly bound to make the instantons. This is because the instantons are still very flimsy and fragile   when  the temperatures $T$ lies inside the window $(\Delta T)/T\leq 1/N$. When the temperature $T$ lies above this window $(\Delta T)/T\gg N^{-1}$ our picture suggests  that $N$ different constituents are strongly correlated to make a well localized instanton with size $\rho\simeq T^{-1}$. We predict that this window must shrink to a point for large $N$. 

$\bullet$ The present work is devoted to a study of 
the conformal window in large $N$ limit when the system consists  $N_f$ quarks in fundamental representations  with    $ \kappa=N_f/ N$ being fixed. This technique can be easily generalized to any other quark representations. Such a generalization  is not a pure academic problem as the corresponding construction  can be  considered as a phenomenologically viable  ``walking technicolor" model, see  recent review on the subject~\cite{Sannino:2009za}. 

$\bullet$  In the present work we advocate an idea that the confinement-deconfinement phase transition in large $N$ limit can be viewed as  $\grave{\rm a}$ la   BKT 
 phase transition. 
 We also explained  (at the end of section \ref{conformal})    why 
 the correlation length $\zeta\sim  \exp(\frac{1}{\sqrt{\kappa_c-\kappa}})$,  which normally accompanies  
 the BKT transition, can not be derived within our approach. 
 We think that the corresponding studies, in principle, can be carried out analytically in the ``deformed QCD" model where one can study the correlation length in the ``plasma phase" side being  in weakly coupled regime \cite{Poppitz:2009uq, Poppitz:2009tw}.  The same hard questions can be also studied   using some numerical approaches\footnote{We would like to make a comment on the terminology which was used in 
  ref. \cite{Faccioli:2013ja}, where the corresponding instanton's constituents were called the ``dyons". This term could be confusing and misleading as the term ``dyon" is normally attributed to a static particle  in $(3+1)$ dimensional  Minkowski space-time, which simultaneously carries  the magnetic and electric charges. The instanton constituents have fundamentally different nature: they are the objects defined in 4d Euclidean space, which  describe the tunnelling events, see footnote \ref{constituents} with more comments and references on the subject.} as advocated recently in ref. \cite{Faccioli:2013ja}. The crucial point is that the partition function in this ``plasma phase" (the confinement  phase in conventional terminology) should be  formulated from the very beginning  in terms of the dominant pseudo-particles, the instanton's constituents. 
  The fact that the the instanton's constituents rather than the instantons themselves are the dominant pseudo-particles in the confinement phase has been advocating by many authors from different perspective, see footnote \ref{constituents} with more comments and references. 
  We leave this subject for future studies.

 $\bullet$The key ingredient of our  approach is the analysis  of the instanton induced potential studied  in the ``molecular side" of the phase diagram where the non-perturbative computations are under theoretical control.
 In pure gauge theory the corresponding induced potential is linked to the $\theta$ dependence (\ref{gamma_N}).
 Therefore, a study of the $\theta$ dependence when the phase transition line is crossed is a key element in understanding of the strongly coupled QCD. If  a single massless quark is present in the system, the $\theta$ parameter can be rotated away   by redefining the quark fields. Such a change of variables does not modify  the physics, of course, as the instanton induced potential remains essentially the same in large $N$ limit. However, a non-vanishing quark mass would produce a drastically  different $\theta$ behaviour  on two sides  of the phase diagram border.  Such drastic changes in $\theta$ behaviour represents   not  a pure   academical problem.  It might  also  lead to some  observational consequences    as the $\theta$ parameter is thought  to be nonzero during the QCD phase transition in early universe.   This unique source of the strong $\cal{CP}$  violation might be a missing ingredient to understand  the dark matter as well as the baryon- antibaryon asymmetry existing in nature, see footnote \ref{CP} and ref.\cite{Lawson:2013bya} with some references and short overview on possible  cosmological consequences  related to a nonvanishing $\theta$ during the QCD phase transition in early universe.
 
 $\bullet$ As we mentioned in the text, one  should not confuse the conventional thermodynamical terms such as ``phase", ``density" or ``entropy" applied to the 4d Euclidean objects  with the corresponding entities which are normally applied to a conventional system of static objects in 3+1 Minkowski space.  However, from the 5 dimensional (holographic) viewpoint    these 4d Euclidean objects can be considered as some  static particles. 
Such a view  on the instantons (instanton quarks) as the static objects from 5d perspective might be worthwhile  for further investigation.

\acknowledgments   
Author thanks Misha Stephanov and Dam Son for a collaboration during the  initial stage of the project.
 I am thankful to Massimo D'Elia, Erich Poppitz and Mirthat Unsal for correspondence and a number of useful comments. 
 This work 
was supported, in part, by the Natural Sciences and Engineering
Research Council of Canada.

\end{document}